\documentstyle[buckow]{article}

\def\ii{{\rm i}}
\newcommand{\eqn}[1]{(\ref{#1})}
\def\beq{\begin{equation}}
\def\eeq{\end{equation}}

\begin {document}
\hfill{LPTENS-02/04}
\def\titleline{
Remarks on the Calculations of Charged Open String Amplitudes: 
the 1-loop Tadpole 
}
\def\authors{
Chong-Sun Chu\1ad , Rodolfo Russo\2ad and Stefano Sciuto\3ad }
\def\addresses{
\1ad
Center for Particle Theory,
Department of Mathematical Sciences, \\
University of Durham, DH1 3LE, UK \\
{\tt chong-sun.chu@durham.ac.uk} \\
\2ad
Laboratoire de Physique Th\'eorique de
l'Ecole Normale Sup\'erieure,  \\
24 rue Lhomond, {}F-75231 Paris Cedex 05, France \\
{\tt rodolfo.russo@lpt.ens.fr}\\
\3ad
Dipartimento di Fisica Teorica, Universit\`a di Torino;\\
I.N.F.N., Sezione di Torino, Via P. Giuria 1,
I-10125 Torino, Italy \\
{\tt sciuto@to.infn.it}
}
\def\abstracttext{
In string theory, there are various physical situations where the
world-sheet fields have a shifted moding. For instance, this is the
case for the twisted closed string in $Z_N$ orbifold or for the
charged open string in a constant electro-magnetic field. Because of
this feature, it is quite challenging to give explicit formulae
describing the string interaction, even for the bosonic case. In this
note, we focus on the case of the charged open bosonic string and
construct the 1-loop tadpole which is an object generating all
$1$-point functions from the annulus in the presence of an external
field. In the operator formalism, this represents one of the basic
building blocks for the construction of a general loop amplitude.
}
\large
\makefront

\section{Introduction}

In the last years we have learned that string theory
has an even richer
structures than was expected. Many interesting phenomena were discovered during
the second string revolution and the concept of {\em duality} has become a
central guideline in all the recent developments. Most of these new features of
string theory are not manifest in the perturbative expansion and this is the
main reason why they have been overlooked in previous analysis. However, even
in this new context, well known techniques of string perturbation theory
proved to be surprisingly useful. For instance, perturbative computations
often represent a first check for conjectures based on expected duality
relations between different theories. Even more important, open string
perturbation theory acquired a new relevance. In general, we have learned that
the boundary conditions for open strings play a central role in many different
contexts: for instance, they give a quantitative description of various
aspects of D-brane physics~\cite{Polchinski:1995mt} and provide a
natural setup for non--commutative theories~\cite{Seiberg:1999vs}.

In the light of these new ideas we reconsider the operator
formalism for the construction of string amplitudes and, in
particular, the old technique of the 
``sewing procedure''. It has already proven to be very
useful in the study of open string theory in the presence of a
constant NS-NS $B$-field; in fact, by using this formalism, we were
able to construct multiloop amplitudes among neutral
strings~\cite{Chu:2000wp},
and had shown that they
are directly related to the Feynman diagrams
of non--commutative field theories. Our goal is to extend this
approach to the case of the charged string, where the boundary
conditions are different on the various borders of the worldsheet. It
is known that this computation is physically relevant in different
contexts, such as the study of open strings in an external
electro-magnetic field~\cite{Bachas:1992bh} or the analysis of moving
D-branes (which is a T-dual version of the previous
system)~\cite{Bachas:1996kx,Billo:1997eg}, or even the study of closed
string amplitudes in $Z_N$ orbifold theories, where the string
coordinates have a similar shifted moding. Here we consider the simple
case of bosonic string theory and derive the expression of the basic
building blocks that in principle can be sewn together to construct
more complicated amplitudes.

In this note, we apply the approach of
Refs.~\cite{Gross:1970xb,Cristofano:1988hb} to the case of charged
bosonic string and construct an object generating all $1$-point
functions from the annulus in the presence of an external field (the
charged tadpole). Our main result is the derivation of this building
block both in the open channel (where the world--sheet looks like an
annulus) and in the closed one (where the surface looks like a
cylinder). We finally comment about the possible use of these tadpoles
in the construction of higher loop charged string amplitudes.

\section{Charged tadpoles}

In the operator formalism, the string interaction is usually described by
means of vertex operators, like $V$ in Eq.~\eqn{eq:3vert}, encoding the
quantum number of the external emitted on-shell state and depending on the
Hilbert space of a virtual emitting string. This asymmetry is cumbersome in
multiloop computation and thus, for this purpose, it is better to use the
so--called Reggeon Vertex~\cite{Sciuto:1969vz} $ \;{}_E\langle W|$,
where also the emitted string is described by means of a Hilbert
space. 

\begin{equation}
  \label{eq:3vert}
V = \; : c(1)\;{\rm e}^{\ii k X(1)}: ~~ \longrightarrow ~~ {}_E\langle W| =
\int  dp \;{}_E\langle p; 0| : {\rm e}^{\left\{-\frac{1}{2\alpha'}
\oint_0 dz X(1-z)\partial_z X^E(z) + \,(ghost) \right\}}:~.
\end{equation}

One can think of the Reggeon Vertex as an "off--shell" generalization of the
usual vertex operator. However, here off--shell does not have the usual
meaning as in field theory. On the contrary, off--shell just means that the
external states have not been specified yet, and thus $ \;{}_E\langle W|$ can
be seen as the generator of all possible three string interaction when it is
saturated with physical (on--shell) states. For instance, the 3-tachyon
interaction is proportional to $\langle k_1 | \;{}_E\langle W | k_2
\rangle_E\,|k_3 \rangle$. In this respect the Reggeon vertex is conceptually on
the same footing as the boundary states which generates all the 1-point
functions on a disk and thus the couplings and the classical field of
the corresponding D-brane~\cite{DiVecchia:1997pr}.

The key property of $ \;{}_E\langle W |$ is the BRST invariance. Thus
one can use a BRST invariant propagator to identify couples of legs
and construct objects with loops and more external states. Again these
new objects can be seen as the generators of more complicated string
amplitudes. The simplest example of this sewing procedure is to derive
from $\;{}_E\langle W |$ the 1-loop
tadpole~\cite{Gross:1970xb,Cristofano:1988hb}, that is an object
generating all $1$-point functions from the annulus. This can be done
simply by computing

\begin{equation}
  \label{deri}
  \;{}_E\langle T_O | = {\rm Tr}\left[{}_E\langle W |\;\frac{b_0}{L_0}\right]
  = \int_0^1 {\rm Tr}\left[{}_E\langle W |\;b_0\,k^{L_0} \right]
  \frac{dk}{k}~,
\end{equation}
and the result is
\begin{eqnarray}
  \label{Tunc}
  {}_{E} \langle T_O | & = & \int\frac{dk}{k^2} \prod_{n=1}^\infty
  (1-k^n)^{2-D}
  \int\frac{d^Dp}{(2\pi)^D}~ {}_{E}\langle 0;0|\; \exp{\left\{\sqrt{2
  \alpha'} \sum_{n=1}^\infty \frac{a^E_n}{\sqrt{n}}\, p\right\}} \nonumber\\ &&
  \exp{\left\{-\frac{1}{2} \sum_{r,s=1}^\infty a^E_s \left[\sum_{m\not{=} 0}
   D_{sr} (\Gamma \rho\, k^m\rho)\right] a^E_r\right\}} ~~{\rm e}^{\alpha'
  p^2 \ln k} \times (ghost)~.
\end{eqnarray}
In the above formula $\rho: z \rightarrow (1-z)$ and $\Gamma: z \rightarrow 1/z$ are projective
transformations, 
$a_n$ are the string modes satisfying the harmonic oscillator algebra,
and, finally, the $D$-matrix form an infinite dimensional
representation of the projective group of weight
zero~\cite{DiVecchia:1989cy}: $D_{nm}(\gamma)= 
\left(\sqrt{m/n}\right)\,1/m!\;\partial_z^m\gamma^n(z) |_{z=0}$, with
$n,m\geq 1$ and $D_{n0}(\gamma)= 1/\sqrt{n}\; \gamma^n(0)$. From now
on we will omit any reference to the ghost part and focus on the
$X$-contribution, where all the novelties related to the external
field are concentrated.

It is possible to derive the same tadpole of Eq.~\eqn{Tunc} by means of a
different sewing procedure. One can start with a slight modification of the
bosonic boundary state
\begin{eqnarray}
  \label{1c1o}
  & \langle 1,1| & =  \int \frac{d^\perp \ell \,d^\parallel p}{(2\pi)^d}\,
  {}_{cl}\langle
  p, \ell;0,\tilde{0}| \;{}_{op}\langle p ,0;0|
  \exp{\left\{ -\sum_{n=1}^\infty a_n\cdot S\cdot \tilde{a}_n
  \right\}} \; {\rm e}^{i \ell\cdot Y} \times \\ \nonumber &&
  \exp{\left\{-\sqrt{2\alpha'}\sum_{n=1}^\infty
  \frac{a_n^{op}}{\sqrt{n}}\cdot p
  -\!\!\!\sum_{n=1,m=0}^\infty\!\!\!  \Big(
  \tilde{a}_n \cdot S D_{nm}(\rho) \cdot a_m^{op} +
 a_n \cdot D_{nm}(\Gamma \rho)   \cdot a_m^{op}\Big) \right\}} ~,
\end{eqnarray}
which describes the interaction of one open and one closed string on a
disk. For completeness Eq.~\eqn{1c1o} is written with mixed Dirichlet
($\perp$) and Neumann ($\parallel$) boundary conditions, even if in
the following we will need the expression with only Neumann
directions, that is $S_{\mu\nu} = \eta_{\mu\nu}$. In general,
$S_{\mu\nu}$ depends on the  boundary conditions one wants to impose
on the border of the disk and here we follow the convention
of~\cite{DiVecchia:1997pr}. In order to construct the
tadpole~\eqn{Tunc} starting from~\eqn{1c1o}, one must propagate the
closed state by means of the usual closed string propagator and
finally saturate it with another boundary state so that a cylinder of
finite length is formed.
The result of this computation ($\int\! d^2 r\, \langle 1,1|b_0 r^{L_0}\;
\tilde{b}_0 \bar{r}^{\tilde{L}_0} | B,S=\eta\rangle $) is
\begin{equation}
  \label{TuncC}
  {}_{E} \langle T_C | = \int\frac{dq}{q^2} \prod_{n=1}^\infty
  (1-q^n)^{2-D}
  {}_{E}\langle 0;0|\;
%
  \exp{\left\{-\frac{1}{2} \sum_{r,s=1}^\infty a^E_s \left[\sum_{m\not{=} 0}
   D_{sr} (\Gamma \rho\, q^m\rho)\right] a^E_r\right\}} ~.
\end{equation}
Formally the only difference
from Eq.~\eqn{Tunc} is the absence of the
Gaussian integration. In fact, by construction, the result \eqn{TuncC}
is written in the closed string channel where no momentum is
exchanged when only Neumann directions are considered. Of course also
the resulting parameterization of the surface is different and one has
the usual relation $|r|^2 = q =\exp{(1/\ln k)}$.

Let us make a remark about the relation of~\eqn{Tunc} and~\eqn{TuncC}.
In general one may wonder whether the tadpole written in the open
channel and the one written in the closed channel are related by the
usual 1-loop ``modular'' transformation. For this purpose, it is
useful to rewrite the infinite products and sums in the above 
formulae~\eqn{Tunc} and~\eqn{TuncC} in terms of the $\theta$-function
$\theta_{11}$ and its derivatives. This is more easily done in the
formulation of Eq.~\eqn{TchC2}, where the exponential is written in
terms of contour integrals. Since the properties of the
$\theta$-function are well-known, the modular transformation can now
be performed explicitly and one finds that, by transforming the
tadpole $\langle T_C|$, Eq.~\eqn{Tunc} is almost exactly recovered,
apart from a single factor of $\tau$: $\langle T_O| = \langle T_C| \;
\tau^{-1}$. 
This disagreement is due to  a subtlety in this
check. In fact, as explained in~\cite{Green:1987mn} for the usual case
of string amplitudes, in order to recover the correct results after a
modular transformation, one need use both the BRST invariance (which
ensures the cancellation of cuts in the parameter $k$) and the
on-shell conditions for the external states (and this is necessary to
get the right power of $\tau =\ln{k}$ ). For our tadpoles the first
condition is satisfied by construction, once $D$ is fixed to $26$, as
usual in the bosonic case. However, it is clear that the on-shell
conditions cannot be imposed until the external legs are saturated
with some physical states. Thus the modular transformation, by its
nature, cannot be performed at the level of off-shell objects such as
the tadpoles \eqn{Tunc} and \eqn{TuncC}; and can
be performed only at the level of amplitudes. And at this
level, $\langle T_C|$ and $\langle T_O|$ generate results that
are in perfect agreement with each other under modular
transformation.


An advantage of
working in the closed channel
is that one can very easily
introduce a constant field-strength in the boundary state, just by
using $S = \frac{1+F}{1-F}$. If, on the contrary, the matrix $S$
appearing in $\langle 1,1|$ is kept trivial ($S=\eta$), then the open
string stretching between the two borders satisfies different boundary
conditions at $\sigma=0$ and $\sigma=\pi$ (we usually refer to this
case as the ``charged string''). As is well known, in this
case the modings of the open string coordinates are  shifted
(we use the conventions of~\cite{Abouelsaood:gd}).
If $X$ is regarded as a complex function
of the world-sheet coordinate $z$, cuts are present and they are
responsible for the difference between the boundary conditions on the
two borders.
As we will see, this makes the computation of the tadpole in the open
channel (and thus of all multiloop amplitudes) quite challenging. On
the contrary in the closed channel no modification on the
$X$-expansion is present and the matrix $S$ can be easily
diagonalized, thus yielding only some phases in various steps of the
computation.  For sake of simplicity we think that $F$ is
non--vanishing only along the two directions $x^1,x^2$; in this case,
in the coordinates $x^\pm= \frac{1}{\sqrt{2}} (x^1\pm i x^2)$, $S$ is
diagonal: $S=\{{\rm e}^{2\pi i\epsilon}, {\rm e}^{-2\pi i\epsilon}\}$,
where $F=\tan{\pi \epsilon}$. The result of the charged tadpole in the
closed channel is
\begin{eqnarray}
  \label{TchC}
  \langle T_C, F | & = & \frac{1}{\cos{\pi \epsilon}} \int\frac{dq}{q^2}
   \prod_{n=1}^\infty (1- q^n)^{2-D}
  \frac{\prod_{n=1}^\infty (1- q^n)^{2}}{\prod_{n=1}^\infty (1- {\rm
  e}^{2\pi i\epsilon} q^n) (1 -{\rm e}^{-2\pi i\epsilon} q^n)}~
  \\ \nonumber &\times& \langle 0,0|\,
  \exp{\left\{- \sum_{r,s=1}^\infty a^{-}_s \left[\sum_{m > 0}
   D_{sr} (\Gamma \rho\, q^m\rho) {\rm e}^{-2\pi i m\epsilon}\right]
 a^{+}_r - \ldots \right\}}~,
\end{eqnarray}
where the dots stand for the complex conjugate of the written exponent.
Notice the overall
factor $\cos{\pi\epsilon}$ that is just the rewriting of the usual
``Born-Infeld'' normalization of the boundary state with a nontrivial
$F$~\cite{DiVecchia:2000uf} or velocity~\cite{Billo:1997eg}.

It is interesting to notice that the above tadpole can be rewritten
directly in terms of fields
\begin{eqnarray}
  \label{TchC2}
  \langle T_C, F | & = & \frac{1}{\cos{\pi \epsilon}}
  \int\frac{dq}{q^2} \prod_{n=1}^\infty (1- q^n)^{2-D}~
  \frac{\prod_{n=1}^\infty \Big(1- q^n\Big)^2}{\prod_{n=1}^\infty (1-
  {\rm e}^{2\pi i\epsilon} q^n) (1 -{\rm e}^{-2\pi i\epsilon} q^n)}~
  \langle 0,0|  \\ \nonumber &\times& 
  \exp{\left\{-\frac{1}{2\alpha'}\oint
  dz dw \,
   \partial X^-(z) \sum_{m > 0}
   \left[\ln \left(1-q^m\frac{1-w}{1-z}\right)\right]
   {\rm e}^{-2\pi i m\epsilon} \partial X^+(w) - \ldots\right\}}~,
\end{eqnarray}
where the contour integral over $z$ and $w$ is around zero.

The modular transformation in the case of a non-vanishing $F$ is
clearly more complicated. In fact, the exponent in~\eqn{TchC2}, cannot
be simply related to the usual $\theta$-function because the phase
encoding the external field enters in a non trivial way in the
exponent and the sum of the $log$'s cannot be rewritten anymore as a
$log$ of product. However, there is a natural guess for the form of
this tadpole. In fact, the complications are concentrated in the
exponent, while the measure can be written also in the open channel,
obtaining the known result~\cite{Bachas:1992bh}. The remaining part of
the expression can be constructed by analogy with Eq.~\eqn{TchC2}. In
fact, in the exponent one would expect the appearance of the Green
function appropriate for the string which is propagating in the
loop. So, in the case  of the uncharged tadpole, or the tadpole in
the closed channel the $log$ is present. On the contrary, when the
loop is described in terms of charged open strings, one would expect
to find the hypergeometric functions that characterize their
tree-level Green function. Thus our guess for the charged tadpole in
the open channel is
\begin{eqnarray}
  \label{TchO}
  && \langle T_O, F | =  F 
  \int\frac{dk}{k^2} \;\tau^{-\frac{D}{2}}\prod_{n=1}^\infty (1- k^n)^{2-D}~
  \frac{\tau\; {\rm e}^{\pi \epsilon^2 \tau}}{\sinh{\epsilon\tau}} \frac{
  \prod\limits_{n=1}^\infty \Big(1-
  k^n\Big)^2}{\prod\limits_{n=1}^\infty (1-
    k^{n+\epsilon}) (1 -k^{n-\epsilon})}~~
  \langle 0,0| \\ && \nonumber \times\, 
  \exp{\left\{-\frac{1}{2\alpha'}\sum_{m>0}\oint dz dw\left[
   \partial X^-(z) \frac{1}{\epsilon} \left(\frac{1-w}{1-z} k^m\right)^\epsilon
  {}_{2}F_{1}\left(1,\epsilon;1+\epsilon;k^m \frac{1-w}{1-z}\right)
  \partial X^+(w) \right.\right. }\\ && \nonumber +\, {\left.\left.
  \partial X^+(z) \frac{1}{1-\epsilon} \left(\frac{1-w}{1-z}
  k^m\right)^{(1-\epsilon)}
  {}_{2}F_{1}\left(1,1-\epsilon;2-\epsilon;k^m \frac{1-w}{1-z}\right)
  \partial X^-(w)\right]\right\}}~.
\end{eqnarray}
In the second line, we have used the Green function $\langle X^-(z)
X^+(w) \rangle$ and the hypergeometric function arises from the sum
over the shifted modes $\sum_{n=0}^\infty \frac{1}{n+\epsilon}
\left(\frac{w}{z} \right)^{n+\epsilon}$, while in the last line the
correlator $\langle X^+(z) X^-(w) \rangle$ has been used. We have
performed a few checks on the above formula and it always yielded
consistent results. For instance, in the $\epsilon \to 0$ limit it
reproduces Eq.~\eqn{Tunc}, including the terms coming from the
gaussian integration over $p$. Another check is possible in the
case $\epsilon = 1/2$, where the modular transformation is doable.
In this case the tadpoles of~\eqn{TchO} and that of~\eqn{TchC2} are
consistently mapped into each other under the usual map
$\tau_o\to -1/\tau_c$. Finally the 2-tachyon amplitude constructed
from \eqn{TchC} and~\eqn{TchO}
agree with each other for all
values of $\epsilon$.

\section{Discussion}

Clearly the main problem of~\eqn{TchO} is its complication. In
particular, if one rewrites it in terms of modes, the matrices
contracting the oscillators (analogous to the $D$'s of Eq.~\eqn{Tunc})
do not seem to be representation of the projective group or of any
other group. This is a major obstacle in the multiloop computation
where one has to compute products of these infinite--dimensional
matrices. On the contrary the tadpole in the closed channel displays
the usual structure related to the projective group. Thus, using
Eq.~\eqn{TchC2} as a starting point for the sewing procedure, it is
not difficult to construct higher loops amplitudes which appear
automatically written in the Schottky parameterization, as usual in
this approach. In particular the results are analytic in the Schottky
multipliers $q_\alpha$, which are the multiloop generalization of the
parameter $q$ describing the length of the cylinder in
Eq.~\eqn{TuncC}. This has a clear physical explanation. In fact, the
poles of the string amplitudes (when they are regarded as functions of
the $q_\alpha$'s) are, by unitarity, related to the mass of the
particles exchanged in the various propagators. In the closed channel
these are the masses of the closed states that are not modified by the
presence of the external electro--magnetic field. On the contrary, in
the open channel one expects some non-analytic behaviour, because the
Virasoro constraint $L_0$ and thus mass-shell condition for the open
strings are deformed by $F$. This feature is already present at the
level of the 1-loop partition function~\cite{Bachas:1992bh}, where
only the mass spectrum of the theory is relevant. The tadpoles
presented here are a starting point towards a multiloop generalization
of the above result. In the multiloop case also the interaction among
charged and uncharged strings must play a role. From the above
considerations, it is clear that, even if the goal is to compute the
multiloop interaction of the charged string in the open channel, it is
easier to use the building blocks written in closed variables to
construct the amplitude and then do the modular transformation on the
final result in order to rewrite it in the desired channel. We will
pursue this approach in a forthcoming paper.

\smallskip
\noindent
{\large \bf Acknowledgments}

\smallskip
\noindent

This work has been presented at the RTN meeting ``The Quantum
Structure of Spacetime and the Geometric Nature of Fundamental
Interactions'' and supported by the European Union under RTN contract
HPRN-CT-2000-00131.


\end{document}